\documentclass[conference]{IEEEtran}

\usepackage{amssymb}
\usepackage{amscd}
\usepackage{amsmath}
\usepackage{graphicx}
\usepackage{epsfig}
\usepackage{amsthm}
\usepackage{graphics}
\usepackage{psfrag}
\usepackage{rotating}
\usepackage{amsmath} 
\usepackage{amsfonts}
\usepackage{url}
\usepackage{color}
\usepackage{epstopdf}
\usepackage{amsthm}
\usepackage{tikz}
\usepackage{mathtools}
\usepackage{multirow}
\usepackage[final]{pdfpages}
\usepackage{enumitem}
\usepackage{multicol}
\usepackage[nolist]{acronym}
\usepackage[font= small]{caption}
\usepackage{subcaption}
\usepackage[ruled]{algorithm2e}
\usepackage{ragged2e}
\usepackage{placeins}
\usepackage{soul}
\usepackage{array}
\usepackage{multirow}
\usepackage{pgfplots}
\usepackage{pgfplotstable}
\usepackage{calrsfs}
\pgfplotsset{compat=1.12}
\usepgfplotslibrary{polar}

\DeclareMathAlphabet{\pazocal}{OMS}{zplm}{m}{n}

\newfont{\bbb}{msbm10 scaled 700}

\newfont{\bb}{msbm10 scaled 1100}
\newcommand{\CC}{\mbox{\bb C}}

\newcommand{\av}{{\bf a}}
\newcommand{\bv}{{\bf b}}

\newcommand{\fv}{{\bf f}}

\newcommand{\hv}{{\bf h}}

\newcommand{\uv}{{\bf u}}

\newcommand{\xv}{{\bf x}}
\newcommand{\yv}{{\bf y}}

\newcommand{\Fm}{{\bf F}}

\newcommand{\Hm}{{\bf H}}

\newcommand{\Um}{{\bf U}}
\newcommand{\Wm}{{\bf W}}

\newcommand{\Xm}{{\bf X}}
\newcommand{\Ym}{{\bf Y}}

\newcommand{\zetav}{\hbox{\boldmath$\zeta$}}

\newcommand{\herm}{{\sf H}}

\newcommand{\transp}{{\sf T}}

\newcommand{\Na}{N_{\rm a}}
\newcommand{\Nrf}{N_{\rm rf}}

\newcommand{\Ns}{N_{\rm s}}

\newcommand{\gtx}{g_{\rm tx}}
\newcommand{\grx}{g_{\rm rx}}

\newcommand{\stxv}{{\bf{s}}_{\rm tx}}

\usepackage{hyperref}
\hypersetup{
    colorlinks=true,
    linkcolor=blue,
    filecolor=magenta,      
    urlcolor=blue,
    }
\makeatletter
\def\@tempa#1{\@xp\@tempb\meaning#1\@nil#1}
\def\@tempb#1>#2#3 #4\@nil#5{%
  \@xp\ifx\csname#3\endcsname\mathaccent
    \@tempc#4?"7777\@nil#5%
  \else
    \PackageWarningNoLine{amsmath}{%
      Unable to redefine math accent \string#5}%
  \fi
}
\def\@tempc#1"#2#3#4#5#6\@nil#7{%
  \chardef\@tempd="#3\relax\set@mathaccent\@tempd{#7}{#2}{#4#5}}

\@tempa\widehat
\makeatother
\begin{document}

\title{Integrated Sensing and Communication with MOCZ Waveform}



	\author{\IEEEauthorblockN{
			Saeid K. Dehkordi\IEEEauthorrefmark{1},
			Peter Jung\IEEEauthorrefmark{2}\IEEEauthorrefmark{1}, 
                Philipp Walk\IEEEauthorrefmark{2}\IEEEauthorrefmark{1},
                Dennis Wieruch\IEEEauthorrefmark{2}\IEEEauthorrefmark{1}, 
                Kai Heuermann\IEEEauthorrefmark{2}\IEEEauthorrefmark{1}, 
			Giuseppe Caire\IEEEauthorrefmark{1}}
			
		    \IEEEauthorblockA{\IEEEauthorrefmark{1}Technical University of Berlin, Germany\\
			\IEEEauthorrefmark{2}MOXZ, Berlin, Germany\\
			Emails: \{s.khalilidehkordi,peter.jung, caire\}@tu-berlin.de, \{philipp, dennis, kai\}@moxz.tech}}
	
	\maketitle

\begin{acronym}
	\acro{AWGN}{additive white Gaussian noise}
	\acro{MIMO}{Multiple-Input Multiple-Output}
	\acro{OTFS}{Orthogonal Time Frequency Space}
	\acro{ISAC}{Integrated Sensing and Communication}
	\acro{sTHZ}{sub-THz}
	\acro{SNR}{Signal-to-Noise Ratio}
	\acro{mmWave}{Millimeter wave}
	\acro{ML}{Maximum Likelihood}
	\acro{V2X}{vehicle-to-everything}
	\acro{OFDM}{Orthogonal Frequency Division Multiplexing}
	\acro{FMCW}{Frequency Modulated Continuous Wave}
	\acro{LoS}{Line-of-Sight}
	\acro{ISFFT}{Inverse Symplectic Finite Fourier Transform}
	\acro{SFFT}{Symplectic Finite Fourier Transform}
	\acro{HPBW}{half-power beamwidth}
	\acro{ULA}{Uniform Linear Array}
	\acro{CRLB}{Cram\'er-Rao Lower Bound}
	\acro{RF}{Radio Frequency}
	\acro{BF}{beamforming}
	\acro{RMSE}{root MSE}
	\acro{AoA}{Angle of Arrival}
        \acro{AoD}{Angle of Departure}
	\acro{ISI}{Inter-Symbol Interference}
	\acro{SI}{self-interference}
	\acro{TDD}{Time Division duplex}
	\acro{Tx}{transmitter}
	\acro{Rx}{receiver}
	\acro{SIC}{Successive Interference Cancellation}
	\acro{PD}{probability of detection}
	\acro{HDA}{Hybrid Digital-Analog}
	\acro{PSD}{Power Spectral Density}
	\acro{FWHM}{full width at half maximum}
	\acro{SLL}{side lobe level}
	\acro{BS}{Base Station}
        \acro{FoV}{Field of View}
        \acro{CFAR}{Constant False Alarm Rate}
        \acro{OS-CFAR}{Ordered Statistic Constant False Alarm Rate}
        \acro{PSLR}{Peak-to-Sidelobe Ratio}
        \acro{DFT}{Discrete Fourier Transform}
        \acro{ISTA}{Iterative Shrinkage Thresholding Algorithm}
        \acro{JHi-FISTA}{Joint Hierarchical - Fast Iterative Soft Shrinkage Thresholding Algorithm}
        \acro{FISTA}{Fast-Iterative Shrinkage Thresholding Algorithm}
        \acro{LISTA}{Learned Iterative Shrinkage Thresholding Algorithm}
        \acro{MMV}{Multiple Measurement Vector}
        \acro{SMV}{Single Measurement Vector}
        \acro{AMP}{Approximate Message Passing}
        \acro{LAMP}{Learned Approximate Message Passing}
        \acro{MSP}{multi-scatter-point}
        \acro{MOCZ}{Modulation on Conjugate-reciprocal Zeros}
        \acro{AF}{Ambiguity Function}
        \acro{UE}{User Equipment}
        \acro{ISL}{Integrated Side-Lobe Level}
        \acro{PSL}{Peak Side-Lobe Level}
        \acro{CAV}{Connected Automated Vehicle}
        \acro{DSRC}{Dedicated Short-Range Communication}
        \acro{V2X}{Vehicle-to-Everything}
        \acro{V2V}{Vehicle-to-Vehicle}
        \acro{LFM}{Linear Frequency Modulation}
        \acro{SC} {Single Carrier}
        \acro{STF}{Short Training Field}
        \acro{CEF}{Channel Estimation Field}
        \acro{HDA-AF}{Hybrid Digital-Analog Ambiguity Function}
        \acro{PA}{Phased Array}
        \acro{CPI}{Coherent Processing Interval}
        \acro{LoS}{Line-of-Sight}
        \acro{IU}{Intended User}
        \acro{MOCZ}{Modulation on Conjugate-reciprocal Zeros}
        \acro{BMOCZ}{Binary MOCZ}

\end{acronym}

\maketitle

%
%
\begin{abstract}
In this work, we propose a waveform based on \ac{MOCZ} originally proposed for short-packet communications in \cite{MOCZ_princ}, as a new \ac{ISAC} waveform. Having previously established the key advantages of MOCZ for noncoherent and sporadic communication, here we leverage the optimal auto-correlation property of Binary MOCZ (BMOCZ) for sensing applications. Due to this property, which eliminates the need for separate communication and radar-centric waveforms, we propose a new frame structure for \ac{ISAC}, where pilot sequences and preambles become obsolete and are completely removed from the frame. As a result, the data rate can be significantly improved. Aimed at (hardware-) cost-effective radar-sensing applications, we consider a \ac{HDA} beamforming architecture for data transmission and radar sensing. We demonstrate via extensive simulations, that a communication data rate, significantly higher than existing standards can be achieved, while simultaneously achieving sensing performance comparable to state-of-the-art sensing systems. 
\end{abstract}
%
%

\section{Introduction}
In the context of upcoming 6G wireless systems, \ac{ISAC} has emerged as one of the key components \cite{ISAC_Survey}. To this end, concepts such as the \ac{CAV} and \ac{V2X}/\ac{V2V} communication have recently attracted considerable attention. Traditionally, automotive radars using specifically \textit{sensing-centric} waveforms have been employed to obtain high-resolution sensing in the mmWave band \cite{Hasch}. These waveforms include \textit{chirp-sequence} (FMCW) radars and more recently, Phase Modulated Continuous Wave (PMCW)-based systems. On the other hand, vehicular communications have traditionally been used to an extent of vehicles exchanging information which may include safety-related information, cooperative driving information for applications such as platooning  or simply to transmit sensory data. To achieve this, current vehicular communication uses the \ac{DSRC} standard which is limited to a maximum data rate of 27 Mbps \cite{DSRC}. The extremely rapid development of automated driving in recent years, mainly due to numerous high-resolution sensors mounted on vehicles, will inevitably lead to huge amounts of data being produced to perform environmental perception. The information from these can be used to create  high-definition evolving maps of the surrounding environment, as well as other use cases which include object classification, trajectory estimation, and reactive measures by the vehicles. It is clear that effectively communicating such quantities of information  using the \ac{DSRC} standard will not be feasible. 

In \cite{kumari2018ieee}, the authors proposed to use the IEEE 802.11ad standard for vehicular \ac{ISAC}. This standard however contains long preambles in the form of \textit{Golay} sequences, which cannot be used for data communication. In such standards, increasing the estimation performance can be achieved as a tradeoff with communication performance \cite{kumari2018ieee}. In contrast, in this work, we present a \ac{ISAC} framework, based on the \ac{MOCZ} waveform \cite{MOCZ_princ,MOCZ_prac}, which simultaneously utilizes the entire packet for sensing and communication. Fig.~\ref{fig:frame_strc} depicts the proposed data frames compared to the IEEE 802.11ad standard. Note that the proposed frame structure is specifically designed to cope with the short coherence times of highly dynamic channels. Moreover, we combine the \ac{MOCZ} signaling with a hardware and energy-efficient \ac{HDA} architecture, to perform high data rate and robust-to-multipath communication as well as medium to long-range sensing in the delay-Doppler-angle dimensions, especially suitable for the high mobility automotive traffic scenarios. 
\begin{figure}[h]
\centering
\hspace{-1.4cm}
\scalebox{.5}{\usetikzlibrary{arrows.meta,shapes.arrows,chains,decorations.pathreplacing}

\newlength\myht
\settoheight{\myht}{$n-2$}
\tikzset{%
  MyStyle/.style={draw, text width=100pt, text height=10pt, text centered,minimum height=\myht+2*3*1mm),fill=green!20},
  MyStylef/.style={draw, text width=60pt, text height=10pt, text centered,minimum height=\myht+2*3*1mm)},
  MyStylefxn/.style={draw, text width=10pt, text height=10pt, text centered,minimum height=\myht+2*3*1mm),fill=yellow!20},
  MyStyleD/.style={draw, text width=100pt, text height=10pt, text centered,minimum height=\myht+3*2*1mm),fill=red!20},
  MyStyleDx/.style={draw, text width=25pt, text height=10pt, text centered,minimum height=\myht+3*2*1mm),fill=red!20},
  MyStyleE/.style={draw, text width=15pt, text height=10pt, text centered,minimum height=\myht+2*3*1mm),fill=yellow!20},
  MyStyleS/.style={draw, text width=40pt, text height=10pt, text centered,minimum height=\myht+2*3*1mm),,fill=red!20},
  boxy/.style = {rectangle,draw=black, minimum height=\myht+2*3*1mm ,fill=blue!20},
  boxy2/.style = {rectangle,draw=black, minimum height=\myht+2*3*1mm ,fill=red!20},
  myarrow/.style={shape=single arrow, rotate=90, inner sep=5pt, outer sep=0pt, single arrow head extend=0pt, minimum height=7.5pt, text width=0pt, draw=blue!50, fill=blue!25}
}

\begin{tikzpicture}[-{Stealth[length=2.5pt]}]

    \begin{scope} [shift={(-4,-2.5)},name=scope1,start chain, node distance=-.5pt]
    \foreach \name [count=\xi] in {1}{ \node[MyStyle, on chain] (vlosI\xi) {STF};}
     \foreach \name [count=\xi] in {1}{ \node[MyStyle, on chain] (vlosfI\xi) {CEF};}
     \foreach \name [count=\xi] in {1}{ \node[MyStylef, on chain] (vlosgI\xi) {Header};}
     \foreach \name [count=\xi] in {1}{ \node[MyStyleD, on chain] (vlosDI\xi) {Data Blocks (BLK)};}
     \foreach \name [count=\xi] in {1}{ \node[boxy, on chain] (vlosEI\xi) {OS};}
  \end{scope}

      \draw [decorate,decoration={brace,amplitude=10pt,mirror}]
  (vlosI1.south west) -- (vlosfI1.south east) node[black,midway,below=8pt]
  {Preamble};

  \begin{scope} [shift={(-4.0,-5)}, start chain, node distance=-.5pt]
    \foreach \name [count=\xi] in {2}{ \node[MyStyleDx, on chain] (vlosfS\xi) {BLK};}
    \foreach \name [count=\xi] in {2}{ \node[MyStyleDx, on chain] (vlosfS1\xi) {...};}
     \foreach \name [count=\xi] in {2}{ \node[MyStyleDx, on chain] (vlosDS\xi) {BLK};}
     \foreach \name [count=\xi] in {2}{ \node[MyStylefxn, on chain] (vlosFxn\xi) {};}
     
     \foreach \name [count=\xi] in {2}{ \node[MyStyleDx, on chain] (vlosf2S\xi) {BLK};}
     \foreach \name [count=\xi] in {2}{ \node[MyStyleDx, on chain] (vlosf2S1\xi) {...};}
     \foreach \name [count=\xi] in {2}{ \node[MyStyleDx, on chain] (vlosD2S\xi) {BLK};}
     \foreach \name [count=\xi] in {2}{ \node[MyStylefxn, on chain] (vlosFxn2\xi) {};}
     
    \foreach \name [count=\xi] in {2}{ \node[MyStyleDx, on chain] (vlosf3S\xi) {BLK};}
    \foreach \name [count=\xi] in {2}{ \node[MyStyleDx, on chain] (vlosf3S1\xi) {...};}
     \foreach \name [count=\xi] in {2}{ \node[MyStyleDx, on chain] (vlosD3S\xi) {BLK};}
  \end{scope}
    \node[right = 0.25cm of vlosD3S1, xshift=0.0cm]{\LARGE $\cdot\cdots$};
    
      \draw [decorate,decoration={brace,amplitude=10pt,mirror}]
  (vlosfS1.south west) -- (vlosDS1.south east) node[black,midway,below=12pt]
  {$\Omega_1$: T$_{CPI}$};

    \draw [decorate,decoration={brace,amplitude=8pt,mirror}]
  (vlosDS1.south east) -- (vlosf2S1.south west) node[black,midway,below=10pt]
  {T$_{swc}$};

    \draw [decorate,decoration={brace,amplitude=10pt,mirror}]
  (vlosf2S1.south west) -- (vlosD2S1.south east) node[black,midway,below=12pt]
  {$\Omega_2$: T$_{CPI}$};

    \draw [decorate,decoration={brace,amplitude=8pt,mirror}]
  (vlosD2S1.south east) -- (vlosf3S1.south west) node[black,midway,below=10pt]
  {T$_{swc}$};

    \draw [decorate,decoration={brace,amplitude=10pt,mirror}]
  (vlosf3S1.south west) -- (vlosD3S1.south east) node[black,midway,below=12pt]
  {$\Omega_3$: T$_{CPI}$};
  
  \node[] at (-7, -2.5)   (b) {(a)};
  \node[] at (-7, -5)   (a) {(b)};
\end{tikzpicture}}
\caption{Frame structures of the (a) IEEE 802.11 ad (SC PHY) standard  (see \cite{kumari2018ieee}), and (b) the proposed frame for sporadic and ultra-short packet communication. Since we consider a spatiotemporal multiplexing scheme, $T_{\rm swc}$ denotes the switching time to steer the Tx/Rx beams toward spatial segments $\Omega_i$.}
\label{fig:frame_strc}
\end{figure}

\section{MOCZ Waveform}\label{sec:moxz_basics}

The basic idea of \ac{MOCZ} modulation is that information bits are mapped directly to zero patterns of complex-valued polynomials $\Xm(z)=\sum_{n=0}^K x_n z^n$ being the $z$-transforms of the $K+1$ time-domain sample sequences $\xv=(x_0,\dots,x_K)^T$ to be transmitted. A multipath channel, represented by a polynomial $\Hm(z)$, acting in time-domain as a convolution $\hv\ast \xv$ with impulse response $\hv$, adds further zeros (roots) to the signal but does not alter the transmitted zeros. In essence, \ac{MOCZ} is a noncoherent modulation for frequency-selective channels since information can be decoded from the received signal without using pilots and multipath channel estimation. Thus \ac{MOCZ} targets communication scenarios where it is difficult or even impossible to obtain accurate channel information, e.g., due to tight latency constraints or sporadic traffic. In particular, in \cite{MOCZ_princ,MOCZ_prac} it has been shown that zero patterns constructed with conjugated-reciprocal zeros are robust to noise (adding noise polynomials $\Wm(z)$) and can be efficiently decoded at the receiver  by testing the magnitudes $|\Ym(z)|$ of the received polynomial $\Ym(z)=\Hm(z)\Xm(z)+\Wm(z)$ over the zero-pattern (called as DiZeT-decoding in \cite{MOCZ_princ}), i.e., without complex root finding. In \cite{MOCZ_prac} it has been demonstrated that \ac{MOCZ} outperforms other noncoherent schemes, like OFDM-IM, OFDM-DPSK and PPM, in terms of bit error rate and throughput. 

\subsection{Radar Properties of Binary MOCZ}
\begin{figure}
\hspace*{-2em}
\includegraphics[width=1.0\linewidth]{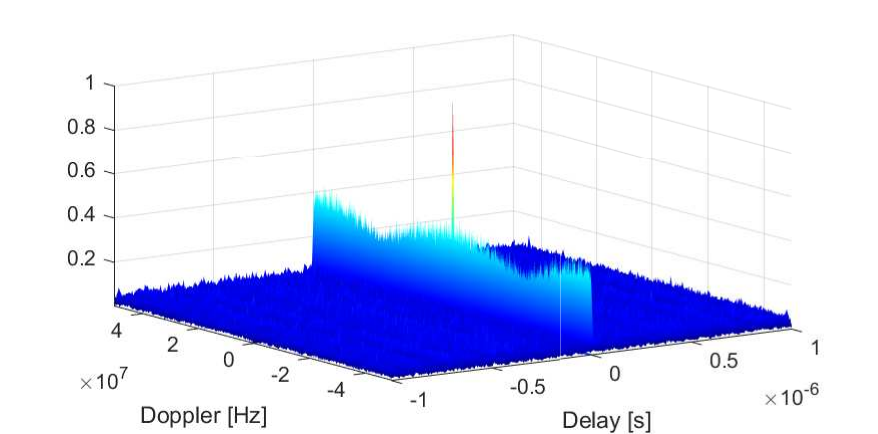}
\caption{Waveform ambiguity function of BMOCZ sequence with length $K=511$.}
\label{fig:ambiguity}
\end{figure}

For ISAC, BMOCZ has an additional appealing feature since the transmit sequences $\xv$ are in this case Huffman sequences having almost-perfect (impulse-like) autocorrelation\footnote{Thus, in this respect BMOCZ is about communication with radar signals.}. For a given bit-sequence $(m_1,\dots,m_K)\in\{0,1\}^K$, the corresponding Huffman polynomial is constructed as:
\begin{align}
\Xm(z) = x_K\prod_{k=1}^{K} (z- \alpha_k)\,~\text{with}~\, \alpha_k=e^{i2\pi\frac{k-1}{K}}R^{2m_k-1}
\end{align}
where the radius is set to $R=\sqrt{1+2\lambda\sin(\pi/K)}$ for $\lambda\approx1/2$ (details see \cite{MOCZ_princ}). For an energy normalized sequence, we set $x_K=-\sqrt{\eta R^{K -2\sum_k m_k}}$ with $\eta=1/(R^K+R^{-K})\in(0,\frac{1}{2})$. It is important to note that all Huffman sequences have the \textit{same autocorrelation}:
\begin{equation}
    \zetav_{xx}\simeq(-\eta,\dots0\dots,1,\dots0\dots,-\eta)\quad
\end{equation}
consisting of the central main peak and only two outer side-peaks, i.e., (one-sided) side-lobe level is $\eta^2$ and peak side-lobe level is $\eta$.
This almost-perfect autocorrelation is illustrated also in the zero-Doppler cut of the 
\ac{AF} in Fig.~\ref{fig:ambiguity} and results in excellent multi-target range estimation and
detection. The `impulsive-like' mainlobe results in a very high resolution and low sidelobe power leads to a high dynamic range.

\section{System model}\label{sec:phy-model}
 The considered \ac{ISAC} system operates over a \ac{mmWave} channel with carrier frequency $f_c$ and bandwidth $W$ sufficiently smaller than $f_c$, such that the narrow-band array response assumptions hold. The  \textit{Fully-Connected} \ac{HDA} \cite{HDA_Sohrabi} host Tx is equipped with $\Nrf$ RF chains driving an antenna array with $\Na$ elements such that  $\Nrf\ll\Na$, and the radar receiver is co-located with the Tx. For simplicity, we assume that the Tx array and the Rx radar array coincide and that the Tx and Rx signals are separated via full-duplex processing \cite{sabharwal2014band}. By letting $\phi\in [-\frac{\pi}{2}, \frac{\pi}{2}]$ denote the steering angle and considering a Uniform Linear Array (ULA) with $\lambda/2$ inter-element spacings, the Tx/Rx array response is given by  $\av(\phi)$, where $\av(\phi)=(a_1(\phi), \dots, a_{N_a}(\phi))^\transp\in \CC^{N_a}$ with
 \begin{align}\label{eq:ULA}
	a_n(\phi)&= e^{j(n-1)\pi \sin(\phi)},\;\; n=1,\dots, N_a .
\end{align}  
\subsection{Transmit/Receive Beamforming and Signal Model}\label{BF_arch}
Let $\Fm\in \CC^{\Na \times \Ns}$ and $\Um\in \CC^{\Na \times \Nrf}$ denote the Tx and the Rx \ac{BF} matrices of the host, respectively. For communication, the host sends $\Ns = Q \geq 1$ data streams through a beamforming matrix $\Fm = [\fv_1, \dots, \fv_{\Ns}]$ where $\fv_q$ denotes the $q$-th column of $\Fm$ associated to the $q$-th data stream. The Tx beamformers $\fv$ are typically chosen such that each covers a relatively narrow (and unique) section of the beam space, and allows a low gain elsewhere, such that $\fv_{q}^\herm\fv_{q'} \approx 0$,~ $ \forall q'\neq q$, i.e., they are approximately orthogonal in the beam space.

By considering a \textit{Spatio-Temporal} multiplexing of the data streams, the Tx operation consists of a single data stream (i.e. $\Fm =\fv$) at any given time and beamforming the Tx stream to other spatial regions in consequent slots. As such, the \ac{Rx} beamforming matrix $\Um\in \CC^{\Na \times \Nrf}$, comprises $\Nrf$ beams that cover the angular region around the \ac{Tx} beam at a given time slot (see Fig.~\ref{fig:channel_scehme}). The Rx \ac{BF} matrices, also known as reduction matrices in the \ac{HDA} context, are selected from a codebook containing a set of $D>\Nrf$ orthogonal beams as $\mathcal{U}_{\text{DFT}} \coloneqq (\uv_{1},...,\uv_{D}) \in \CC^{\Na\times D}$ such that (i.e.$, ~ \uv_{d}^{H}  \uv_{d'}\approx 0 ~ \forall d\neq d'$). In most literature, the $D$ beams are selected from the Fourier basis ($\in \CC^{\Na\times \Na}$). However, other choices such as custom-designed flat-top beams \cite{OTFS_MIMO_AV_22} or multi-taper beams (i.e. Discrete Prolate Spheroidal (Slepian) Sequences \cite{Pedraza}) can be selected. 

The complex baseband continuous-time representation of a single \ac{MOCZ} sequence is given by:
\begin{align}
s(t) = \sum_{n =0}^{K} x[n]\gtx(t-nT) 
\end{align}
where $\gtx$ is the unit energy pulse-shaping filter, and $x[n]=x_n$ are the samples of the single-carrier \ac{MOCZ} transmit sequence $\xv$  such that $\mathbb{E} \left[|x[n]|^2\right] = 1/(K+1)$. The sample period is denoted by $T = 1/W$, where $W$ is the total RF bandwidth and $T_{CPI}$ denotes the \ac{CPI}. Due to the typically small value of $T_{CPI}$, the acceleration and the relative velocity of non-stationary targets such as other vehicles in the environment can be assumed as constants. The transmitted signal at the host vehicle is given by $\stxv(t) = \fv s(t), ~ ~0 \leq t \leq T_{CPI}$,  where $\fv \in \CC^{\Na} $ is the Tx beam-forming vector at the host vehicle.
The communication channel between the host and \ac{IU} vehicle in the time-delay domain is modeled as: 
\begin{align}\label{eq:ComChannel}
\Hm_C (t, \tau) =  \sum_{p=0}^{P-1} h_{p} \bv(\theta_{p}^{c}) \av^\herm (\phi_{p}^{c})\delta(\tau-\tau_{p}^{c}) e^{j(2\pi \nu_{p}^{c})t}\,,
\end{align}
where the $p$th path is defined by  a Doppler shift of $\nu_{p}^{c} = \frac{v^{c}_{p}  f_c}{c}$ at carrier- frequency $f_c$ and Doppler velocity $v^{c}_p$, and $\tau^{c}_p$ is the delay from path length $d_p$. $\bv(\theta^{c})$ is the array steering vector at the \ac{IU}. By assuming an isotropic Rx array at the \ac{IU}, the array response $\bv(\theta^{c}) = 1$. $h_{p}$ is the communication channel gain given by |$h_{p}|^{2} = \kappa_p\frac{\lambda^2 }{(4\pi)^2 (d_p)^2}$, where a path loss exponent of $2$ is considered, which is typical for mmWave LOS outdoor urban and rural scenarios. $\kappa_p$ is the Rician factor and $\psi_p$ a random phase.

The mmWave radar channel for a \ac{CPI} is modeled as doubly time-frequency-selective, which is widely used in automotive radar \cite{Bazzi}. The resultant channel from $Q$ targets results in the $\Na\times\Na$ time-varying MIMO channel with matrix impulse response given by \cite{vitetta2013wireless}
\begin{align}\label{eq:RadChannel}
\Hm_R (t, \tau) =  \sum_{q=0}^{Q-1} \rho_{q} \av(\phi^{r}_{q}) \av^\herm (\phi^{r}_{q})\delta(\tau-\tau^{r}_{q}) e^{j2\pi \nu^{r}_{q} t}\,,
\end{align}
Following the radar equation, the channel coefficient $\rho_{q}$ satisfies $|\rho_{q}|^{2} = \frac{\lambda^{2}\sigma_{{\rm rcs}, q}}{(4\pi)^{3}r^{4}_{q}}$, where for each target,  $\rho_{q}$ is a complex channel gain including the \ac{LoS} pathloss and the radar cross-section coefficient $\sigma_{\mathrm rcs}$. $\nu_{q}^{r} = \frac{2v^{r}_{q}  f_c}{c}$ at Doppler velocity $v^{r}_q$, and $\tau^{r}_q$ is the round-trip delay from path length $r_q$. Over the \ac{CPI}, the channel parameters $\{\rho_{q}, \phi^{r}_{q}, \nu^{r}_{q}, \tau^{r}_{q}\}_{q=1}^Q$ are assumed to remain constant.  Note that multi-path scatterers are also contributing factors to the radar channel, however, \ac{mmWave} channels are characterized by large isotropic attenuation, such that multipath components are typically much weaker than the \ac{LoS} and disappear below the noise floor after reflection, in particular for the backscatter channel seen by the radar receiver (e.g., see \cite{kumari2018ieee,nguyen2017delay}). Note that the LOS path coincides for both channels, however, this does not necessarily hold for other path components. 
 At the output of the radar \ac{Rx} filter-bank, we adopt a generic receive shaping pulse $\grx$. Then by sampling this output, at $t=nT$ we obtain the discrete-time representation of the received $\Nrf$-dimensional signal with length $N$ corresponding to the n-th symbol during a \ac{CPI} as: 
\begin{align}
y[n] =  \sum_{q=0}^{Q-1}\rho_q {\Um}^\herm \av(\phi^r_q) \av^\herm(\phi^r_q) \fv~s(nT - \tau^r_q) e^{j2\pi \nu^r_q nT}  \label{disc_rx_radar_sig}
\end{align}
\section{Radar Detection and Parameter Estimation}\label{sec: Super-Res} 
Due to page restrictions, we refer the readers to \cite{MOCZ_prac,MOCZ_princ} for the detection and channel estimation at the user end and only provide some details on the radar detection and estimation in this section.
The radar target detection problem can be achieved by applying a thresholding function for a specified false alarm probability on the cross-correlation output with the TX signal \eqref{cross_corr}.
We consider an adaptive threshold. Namely, we use the Ordered Statistic - CFAR method, which is known to provide good performance in a realistic scenario when the noise and interference power levels are not homogenous (see e.g. \cite[Chapter 6.5]{richards2014fundamentals}).\\[.5em]
%
{\bf 1) Delay Estimation:}
The (circular-) cross-correlation output between the received
signal and the transmitted sequence is used for radar detection and estimation (as used in classic radar detection). The cross-correlation $\zeta_{xy}[n],~~ n \in [0:N-1]$ in the discrete domain is given by: 
\begin{align}
    \zeta_{xy}[n] = \sum_{m = 0}^{N-1}x^*[m-n]y[m] \label{cross_corr}
\end{align}
For the continuous-valued delay $\tau = \tau_F + \tau_I$, the integer delay $\tau_I$ is then estimated by searching for the lag which maximizes the magnitude of correlation output, i.e. $\tau_I = \underset{n}{\text{arg~max}}|\zeta_{xy}[n]|$~. For estimation of the fractional delay component $\tau_F$, a parabola is fitted at the peak of the cross-correlation.\\[.5em]
%
{\bf 2) Angle Estimation:}
We consider a set of fixed Rx BF vectors where by using a tunable set of phase shifters, the beams are rotated and pointed toward the intended direction. Using this design, prior to delay and Doppler estimation, we calculate the sample covariance matrix from the $N$ sample Rx-signal $\text{cov}(\yv) = \frac{1}{N}\sum_{n=0}^{N-1}\yv[n]\yv^H[n]$. \textit{Beam domain} variant of spectral methods such as MUSIC and ESPRIT can then be used to obtain  high-resolution estimates of the \ac{AoA}.\\[.5em]
%
{\bf 3) Doppler Estimation:}
The Doppler velocity estimation is achieved via frequency synchronization, which uses a correlation-based estimation method over single/multiple frames. Since velocity resolution is proportional to integration time (time on target), we propose a multi-frame linear estimator based on the BLUE\cite{BLUE}. Note that in \cite{kumari2018ieee} (and refs. therein) this is achieved by using the preambles of a very large number of 802.11ad frames, and only a small fraction of these frames comprise data. In our proposed spatio-temporal multiplexing scheme, this is achieved by sending short data packets to distinct segments of the beam space in TDD manner and processing the frames from the same segment jointly. 


\section{Numerical Results}\label{sec:numeric_res}
\begin{figure}
	\centering
	\includegraphics[width=0.95\linewidth]{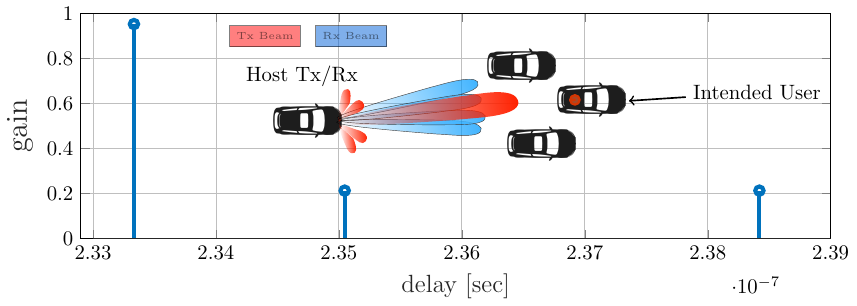}
\caption{Traffic scenario and profile of a modified TDL-D channel.}
\label{fig:channel_scehme}
\end{figure}

\begin{figure}
	\centering
    \hspace*{-1em} 
	\includegraphics[width=0.95\linewidth]{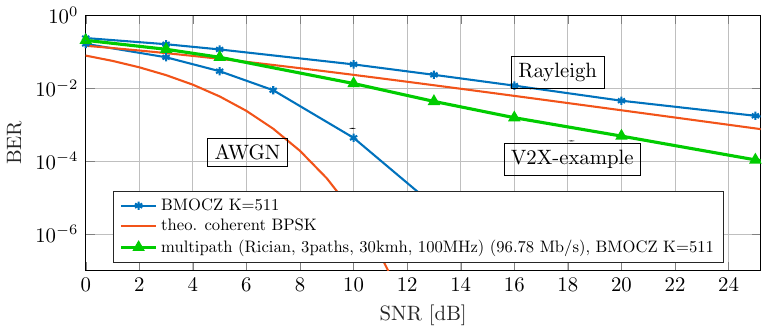}
	\caption{Comparison of BER for BMOCZ ($K=511$) with DiZeT-decoding over various channel models: AWGN, flat-fading Rayleigh and frequency-selective Ricean fading (see Fig. \ref{fig:channel_scehme}). $3$dB-gap wrt. coherent BPSK performance is consistent with results in \cite{MOCZ_princ}.}
	\label{fig:BER_Comp}
\end{figure}

\begin{figure}[t]
\centering
\includegraphics[width=88mm]{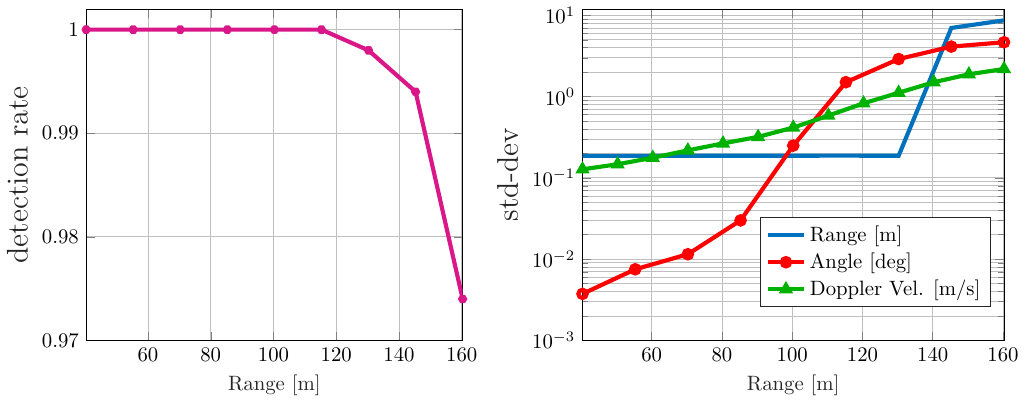}
\caption{Radar parameter estimation performance of MOCZ in the considered traffic scenario.}
\label{fig:rad_est}
\vspace{-\baselineskip}
\end{figure}

We consider the traffic scenario depicted in Fig.~\ref{fig:channel_scehme}, where the adjacent vehicles cause multipath components in the received signal at the user. The values used in the simulation are provided in Tab.~\ref{tab:fontsizes}. The main and multi-path components are modeled as Rician with a factor of 10 as in \eqref{eq:ComChannel}. By using a moderately small BW of $100$ MHz, Fig.~\ref{fig:BER_Comp} shows the BER performance of the depicted scenario for a targeted data rate of approx. $100$ Mb/s at the user with a uncoded payload of $511$ bits. Simultaneously, we use the backscattered data signal to perform radar estimation of the user. The presented curves in Fig.~\ref{fig:rad_est} are averaged over 1000 monte-carlo runs. Note that the system's operational range can be extended by increasing the Tx power (equivalently EIRP).

{
%
\setlength{\tabcolsep}{2mm}%
\renewcommand{\arraystretch}{1.2}
\newcommand{\CPcolumnonewidth}{not used}%
\newcommand{\CPcolumntwowidth}{21mm}%
\newcommand{\CPcolumnthreewidth}{12mm}%
\newcommand{\CPcolumnfourwidth}{33mm}%
\begin{table}[h]
\caption{System Parameters}
\scriptsize
\centering
	\begin{tabular}{|c|c|}
		\hline
		$f_c=60.0$ [GHz] & $W=100$ [MHz] \\ \hline
		$\text{EIRP}=35$ [dBm] & $\sigma_{\mathrm{rcs}}=10$ [dBsm] \\ \hline
		P$_{\rm fa}$ $=1e-4$  & Noise PSD  = $2\cdot10^{-21}$ [W/Hz] \\ \hline
		$\Na=64$~,~$\Nrf=4$& K=511  \\ \hline
	\end{tabular}
\label{tab:fontsizes}
\end{table}
}

\bibliography{IEEEabrv,book}


\end{document}